\documentclass[sigconf]{acmart}

\usepackage{gensymb}
\usepackage{amsmath}
\usepackage{graphicx}

\usepackage{subfig}
\usepackage{tabularx}
\usepackage{booktabs}

\usepackage{balance}

\AtBeginDocument{%
  }

\copyrightyear{2024}
\acmYear{2024}
\setcopyright{acmlicensed}\acmConference[ISWC '24]{Proceedings of the 2024
ACM International Symposium on Wearable Computers}{October 5--9,
2024}{Melbourne, VIC, Australia}
\acmBooktitle{Proceedings of the 2024 ACM International Symposium on
Wearable Computers (ISWC '24), October 5--9, 2024, Melbourne, VIC, Australia}
\acmDOI{10.1145/3675095.3676611}
\acmISBN{979-8-4007-1059-9/24/10}

\begin{document}

\newcommand{\name}{EchoGuide}{}

\newcommand{\saif}[1]{\textcolor{red}{{#1}}}
\newcommand{\delete}[1]{{\sout{ #1}}}





 \title{\name{}: Active Acoustic Guidance for LLM-Based Eating Event Analysis from Egocentric Videos}




\author{Vineet Parikh}
\affiliation{%
  \institution{Cornell University}
  \city{Ithaca, NY}
  \country{USA}}
\email{vap43@cornell.edu}
\orcid{0009-0000-8791-9340}

\author{Saif Mahmud}
\affiliation{%
  \institution{Cornell University}
  \city{Ithaca, NY}
  \country{USA}}
\email{sm2446@cornell.edu}
\orcid{0000-0002-5283-0765}

\author{Devansh Agarwal}
\affiliation{%
  \institution{Cornell University}
  \city{Ithaca, NY}
  \country{USA}}
\email{da398@cornell.edu}
\orcid{0009-0005-1338-9275}


\author{Ke Li}
\affiliation{%
  \institution{Cornell University}
  \city{Ithaca, NY}
  \country{USA}}
\email{kl975@cornell.edu}
\orcid{0000-0002-4208-7904}

\author{François Guimbretière}
\affiliation{%
  \institution{Cornell University}
  \city{Ithaca, NY}
  \country{USA}}
\email{francois@cs.cornell.edu}
\orcid{0000-0002-5510-6799}

\author{Cheng Zhang}
\affiliation{%
  \institution{Cornell University}
  \city{Ithaca, NY}
  \country{USA}}
\email{chengzhang@cornell.edu}
\orcid{0000-0002-5079-5927}

\renewcommand{\shortauthors}{Vineet Parikh et al.}

\begin{abstract}
Self-recording eating behaviors is a step towards a healthy lifestyle recommended by many health professionals. However, the current practice of manually recording eating activities using paper records or smartphone apps is often unsustainable and inaccurate. Smart glasses have emerged as a promising wearable form factor for tracking eating behaviors, but existing systems primarily identify when eating occurs without capturing details of the eating activities (E.g., what is being eaten). In this paper, we present \name{}, an application and system pipeline that leverages low-power active acoustic sensing to guide head-mounted cameras to capture egocentric videos, enabling efficient and detailed analysis of eating activities. By combining active acoustic sensing for eating detection with video captioning models and large-scale language models for retrieval augmentation, \name{} intelligently clips and analyzes videos to create concise, relevant activity records on eating. We evaluated \name{} with 9 participants in naturalistic settings involving eating activities, demonstrating high-quality summarization and significant reductions in video data needed, paving the way for practical, scalable eating activity tracking.

\end{abstract}


\acmArticleType{Research}

\maketitle
\newif\iffvg
\fvgfalse

\section{Introduction}
\label{sec:intro}



%

Self-recording eating behaviors is a step towards a healthy lifestyle recommended by many health professionals. However, the current practice requires users to manually record their eating activities, including when and what they eat, using paper records or smartphone apps. This manual method is often unsustainable and sometimes inaccurate, as users frequently forget to record their activities.

Smart glasses have emerged as a promising wearable form factor for tracking eating behaviors. To alleviate the need for manual recording, various sensing systems based on smart glasses have been developed to distinguish eating behavior from arm movements \cite{eatingtrak}, ambient sound\cite{thomaz2015inferring} or facial muscle movements\cite{mydj}. However, most of these systems can only identify when eating occurs but not what is being eaten, which is critical information for interpreting eating behaviors. Conversely, sensing systems such as cameras, which can capture detailed information on eating (e.g., the type of food consumed), have high power consumption, making continuous operation impractical on commodity smart glasses.

In this paper, we explore the research question:

\begin{itemize}
    \item \textit{Is it possible to use low-power active acoustic sensing on glasses to automatically guide the camera to capture activities, such as eating, in an energy-efficient manner without losing much critical information?}
\end{itemize}

Active acoustic sensing\cite{10.1145/3161188} has been shown as a low-power and powerful sensing modality for tracking and interpret various types of fine-grained body poses on wearables, including facial expressions\cite{eario}, gaze\cite{10.1145/3636534.3649376}, finger pose\cite{yu2024ring,lee2024echowrist}, body pose\cite{PoseSonic}, tongue gesture\cite{10.1145/3594738.3611358},  silent speech recognition\cite{echospeech,zhang2023hpspeech}, authentication\cite{li2024sonicid,10.1145/3643832.3661890} and physiological signal \cite{chen2024exploring,fan2023apg}. The latest work ActSonic\cite{ActSonic} has shown that using active-acoustic sensing on glasses can recognize over 28 types of everyday activities (including eating) in the wild with 89\% F1 score at each second without the need for any training data from a new user. More specifically, it recognizes eating activities with an F1 score of 90\% in completely unconstrained environments. However, this sensing modality doesn't capture the full context of a given activity. For activity recording and downstream applications (such as calorie counting or recipe assistance), it's important to understand not only \textbf{what action} (e.g., when eating happens) is performed given body motion, but also \textbf{what objects} (e.g., what is the food being eaten) the action is being performed with.

In this paper, we present the design and implementation of \name{}, an application and system pipeline that combines the strengths of active acoustic sensing for action detection, video captioning models for detailed egocentric action understanding, and large-scale language models with retrieval augmentation for conversational QA with action records, to enable efficient and seamless action recording and retrieval within specialized everyday domains such as eating. 

By leveraging efficient pre-trained models for action detection via active acoustic sensing from ActSonic\cite{ActSonic}, \name{} can intelligently ``clip" the videos to guide the camera and video models, creating an activity record that remains far shorter than naive dense-clip video captioning applications while additionally remaining far more relevant than inflexible sparse clipping methods.

We evaluate the performance of \name{} with 9 participants wearing GoPros and ultrasonic sensors affixed to commodity eyeglasses to collect data about eating in the unconstrained environment of the participant's choice. With customized acoustic data preprocessing, action detection, video captioning, and action retrieval QA pipeline, we efficiently build activity records with significant reductions in record length while maintaining high semantic correlation with densely captured records. We evaluate the system via a semi-in-the-wild user study with 9 participants focused on eating actions. Additionally, we discuss some of the challenges that \name{} must address to be deployed further at scale. We summarize the contributions as follows:
\begin{itemize}
    \item To the best of our knowledge, we are the first to demonstrate the feasibility of leveraging active acoustic sensing on glass frames to guide the highly efficient capture and analysis of egocentric video for eating activity tracking.
    
    \item We propose an end-to-end application pipeline enabling seamless and efficient action detection, video captioning, and action retrieval/QA leveraging a combination of active acoustic sensing and egocentric video.
    
    \item We evaluated the end-to-end pipeline on eating activities collected in naturalistic settings of 9 participants' choices through a user-independent and session-independent study. Our system provided high-quality summarization (68\% average reduction in activity records along with high alignment between reduced and original activity records given eating-focused videos) while significantly reducing the amount of video data needed.

\end{itemize}

\section{Related Work}
\label{sec:related-work}

\textbf{Multimodal Image/Video Captioning and Summarization: } As larger-scale language and multimodal generative  ``foundation models" \cite{FM} have been trained and released, image and video captioning has extended from simply determining the similarity of images/videos to a premade list of captions \cite{Sembed, Jpose, BEVT, CLIP, XCLIP, Ego-VLP} towards generating captions for new videos based on either fine-tuning inexpensive smaller-scale Large Language Models with video encoders and captions \cite{lavila} or leveraging the emergent properties of natively multimodal Large Language models (such as the OpenAI GPT-4 multimodal series \cite{GPT4, GPT4o} and Google Gemini multimodal series \cite{Gemini}), truly ``open-world" video captioning becomes more possible especially in a ``zero-shot" paradigm without labeled examples. These systems can be applied offline throughout a video to create ``activity records": long documents which encode which activities a person might be completing within the course of a video, and which can be efficiently indexed and searched.

Extracting insights from preprocessed "activity records" requires methods which can generate relevant answers to queries that are grounded in specific documents. Recent generative methods, especially in scenarios involving domain-specific information, leverage the Retrieval-Augmented Generation \cite{RAG} for returning helpful responses given queries and documents containing relevant information.

However, the primary bottleneck when leveraging image/video captioning and summarization systems especially over longer videos is power and compute consumption: wearable cameras such as the GoPro HERO9 \cite{gopro} do not have sufficient battery life for continuous daily capture, and video-processing models which recognize activities and objects have high compute requirements. 

\textbf{Eating Recognition on Glasses: }
Various sensing modalities on eyeglasses form factors have been proposed to track eating events. These modalities include EMG electrodes~\cite{zhang2017monitoring}, piezoelectric sensing~\cite{mydj, farooq2016novel}, contact microphones~\cite{auracle}, microphones and IMUs~\cite{mirtchouk2017}, and sensor fusion~\cite{fitbyte, fitnibble}. While these systems track eating episodes, they lack the ability to provide detailed information related to eating activities, such as what food a person eats. This limitation is due to the absence of optimized access to an egocentric camera for extended monitoring periods.
\section{The System Design of \name{}}
\label{sec:sensing-mechanism}

\begin{figure}[h]
    \centering
     \includegraphics[width=1\linewidth,height=2.5cm]{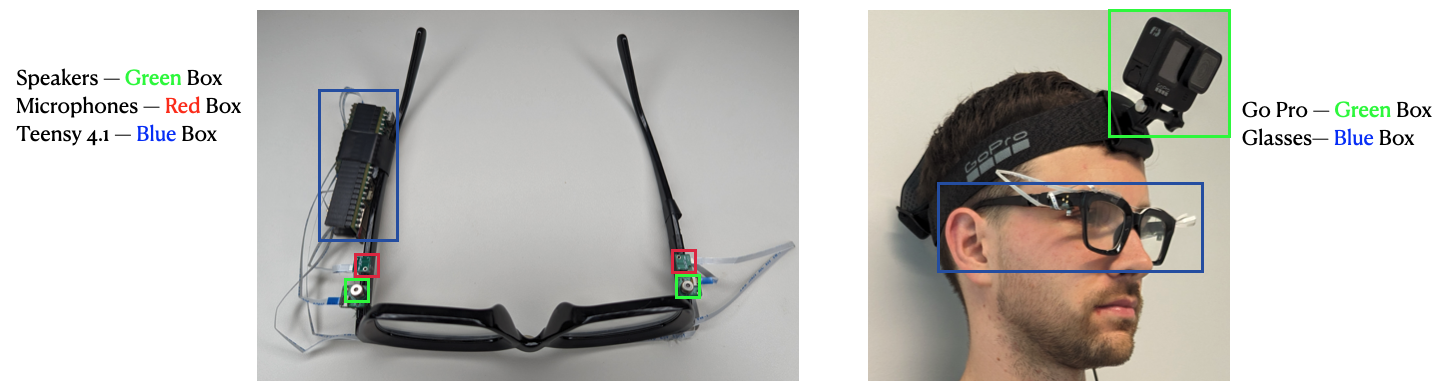}
     \caption{Glasses and GoPro Hardware setup for \name}
     \label{fig:hardware}
 \end{figure}

In this section, we will present the design of \name{} including 1) the hardware prototype we used to collect egocentric acoustic and video data for eating activities; and 2) the software and deep learning pipeline we used to process the acoustic data for eating event segmentation and extract details of eating episodes from the segmented video clips. 


\subsection{Hardware Prototype}
\subsubsection{Glasses with active acoustic Sensing}
We used a similar hardware prototype design of the glasses as ActSonic\cite{ActSonic} as shown in Figure \ref{fig:hardware}. They include two OWR-05049T-38D\footnote{\url{https://www.bitfoic.com/detail/owr05049t38d-14578}} speakers for chirps and two ICS-43434\footnote{\url{https://invensense.tdk.com/products/ics-43434/}} microphones for receiving signals. The system uses a Teensy 4.1\footnote{\url{https://www.pjrc.com/store/teensy41.html}} microcontroller to store the transmitted signal and save the received signal on its SD card. Using a similar hardware prototype design will allow us to directly use ActSonic's fully pre-trained deep learning model to identify eating moments in everyday activities without the need of any new training data.

\subsubsection{Head-mounted GoPro for egocentric video capture}: 
To collect egocentric video of user's activities, we used a head-mounted GoPro HERO-9\footnote{\url{https://gopro.com/en/us/news/hero9-black-announce}}, as shown in Figure \ref{fig:hardware}. The data was saved on the SD card within the GoPro. 

\subsection{Data Processing Pipeline}
\begin{figure}[h]
    \centering
     \includegraphics[width=1\linewidth,]{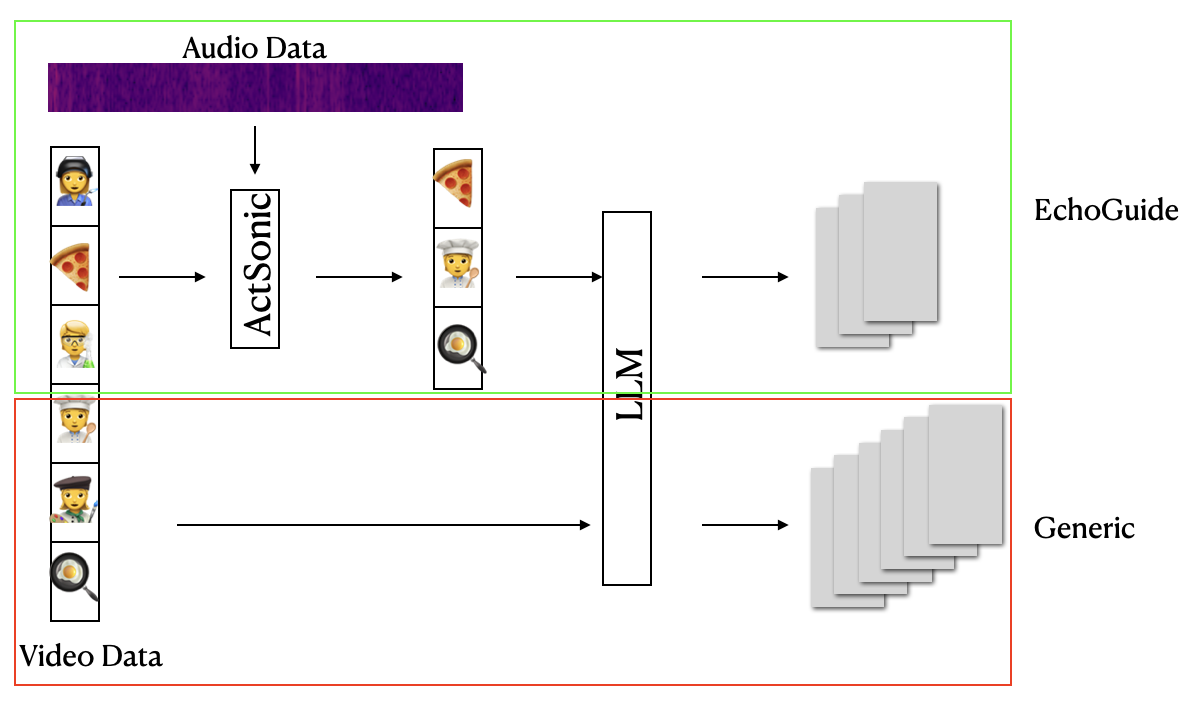}
     \caption{ \name \space vs Generic LLM Document Generation}
     \label{fig:SucrePipeline}
 \end{figure}

\label{sec:dl-model}
The \name{} Software and Deep Learning application follows a two-stage pipeline consisting of first processing synchronized egocentric videos and processed active acoustic data into ``activity records" representing a history of user activities, and then building a question-answering framework leveraging large language models and Retrieval Augmented Generation for indexing, retrieving, and answering questions grounded in these ``activity records". This modular pipeline enables incremental improvement of individual components as they become more capable and is contrasted with a naive dense captioning pipeline which must process the entire video (as shown in Fig \ref{fig:SucrePipeline}), with natural language acting as an intermediate step between dense perceptual information and question-answering systems.

\subsubsection{Using Active Acoustic Sensing to localize relevant actions and clip videos}
We acquired the Resnet-18 model reported in ActSonic\cite{ActSonic} via contacting the authors. Our goal is to directly use the pretrained model in ActSonic to determine when an eating activity happens, leveraging the strong user-independent performance of ActSonic for detecting eating activities in in-the-wild settings \cite{ActSonic}.

We leverage ActSonic's ResNet-18 model as an event detector by splitting the active acoustic differential echo profile (synchronized with the video) into consecutive 2-second windows which can be passed directly into the model. We define a set of ``domain-specific classes" within the label space of ActSonic which capture important events for this particular domain, extract class predictions for all sliding windows (essentially treating the class prediction's ``timestamp" as the last timestamp of the corresponding window), and construct intervals of events by filtering for ``domain-specific classes" and joining equivalent predictions in consecutive windows to create clips without requiring dense captions.

\subsubsection{Generating activity records from video and active acoustic sensing}
To generate activity records, we take a preprocessed and synchronized dataset containing egocentric videos and acoustic echo/differential profiles from user activities, and apply two modules: a \textbf{clipper} to each long untrimmed video/acoustic pair to convert the pair into a series of video clips with possible metadata, and a \textbf{captioner} which can take video clips and associated metadata (e.g. timestamp, acoustic classifier label, etc) and generate a ``caption" for the clip in EGO-4D format (treating "C" as the camera-wearer) \cite{ego4d}, incorporating time metadata as well. We can then join the captions with timestamps to create an ``activity record" for the given session. Within \name(), we primarily focus on proving out the combination of ActSonic as a ``clipper" \cite{ActSonic} and LaViLa's Narrator (a video-to-GPT2 model fine-tuned on EGO-4D \cite{ego4d}, an egocentric vision dataset) as a ``captioner" \cite{lavila}.

\subsubsection{Answering questions given activity records}
We leverage a Retrieval-Augmented Generation \cite{RAG} framework such as LlamaIndex\cite{LlamaIndex}) for efficiently chunking and embedding a given series of documents (leveraging OpenAI's ``text-embedding-ada-002" embedding model) as well as input queries. Given an input query, we run a similarity search on the query embedding vs chunk embeddings (using cosine similarity) and pass the top ``k" chunks into the context of a language model (in our case GPT3.5 \cite{GPT3}) to efficiently answer questions about the activity record via a chat/question-and-answer interface. 

\section{User Study}
\label{sec:user-study}
To collect data for evaluating \name{}, we conduct a semi-in-the-wild user study in various naturalistic locations (including participant homes and offices), focusing on capturing natural data of users eating while also performing other activities (such that only parts of each sequence relate to relevant actions). We leverage the activity set proposed in ActSonic \cite{ActSonic}, which describes a wide collection of everyday activities.

\textbf{Participants} The \name{} user study received approval from the Institutional Review Board for Human Participant Research (IRB) at our organization. We recruited 10 participants for a semi-in-the-wild user study at their homes. However, 1 participant's data was lost during the user study. Therefore, we ended up with 9 valid participants in the study, ranging in age from 19 to 34. 6 participants self-reported as male while 3 self-reported as female. We collected basic demographic data and their ratings on the prototype through an IRB-approved questionnaire. The average comfort rating on a Likert scale of 0 to 5 was 2.62.

\textbf{Data Capture Apparatus} We captured acoustic data using the sensing system integrated into \name{} eyeglasses and recorded egocentric activity video data via the \name{} GoPro Hero9  \cite{gopro} camera mounted on the participants' heads using a lightweight body mount from the same manufacturer. The camera's horizontal and vertical field of view was set to $118\degree$ and $69\degree$ respectively. It recorded egocentric videos at a resolution of $720$p and a frame rate of $30$ fps.

\textbf{Study Procedure} We conducted a 9-participant semi-in-the-wild user study in unconstrained environments such as participants' homes and offices. The recruited participants were equipped with eyeglasses and a head-mounted camera. We synchronized the acoustic and video data with a clap action performed by the researcher as the two sensors were physically separated. After synchronization, the participants could continue normal activities without interruption if they ate or drank at least one item within the 40-minute window. Upon completing the 40-minute study, the participants returned the devices to the researcher. We had 5 participants collect data at their homes and 4 in their office environments. 

\section{Evaluating quality of eating activity summaries and responses with LLMs}
\label{sec:eval}

\subsection{Metrics}
To measure the value of leveraging a supervised ultrasonic model to actively guide video captioners toward more efficient action captioning for retrieval, we define two primary metrics for evaluating system quality:
\subsubsection{Answer alignment with dense captioning}
Given a single question related to the domain, we find semantic similarity between the answer from a RAG QA agent that has indexed an activity record with alternative sampling (e.g. leveraging the ultrasonic modality to filter and caption fewer frames) and the answer from a RAG QA agent that has indexed an activity record with dense video sampling (e.g. captioning the entire video, which can reduce overall efficiency but captures all possible information). Semantic similarity is captured via BERT F-1 scores \cite{BERTScore}, which captures pairwise cosine similarities (within the range -1 to 1) between BERT output embeddings to capture semantic and contextual information and which shows high correlation with human evaluations on summarization and captioning tasks (closely related to this work). Similarity scores are used to quantify information loss between the densely captioned model and the ActSonic-captioned model.

\subsubsection{Recording reduction compared to dense sampling}
 Different video clipping methods can lead to different ``line counts" for an activity record (as each line of an activity record correlates to a clip in the video where a video captioner model was used). We can therefore find the size reduction between \name{}/ActSonic records (where clips are extracted using the ultrasonic modality) vs densely-captioned records (where clips are densely extracted at 1-second intervals).

\subsection{Evaluation Procedure/Baseline Description}
We show per-participant metrics across the two studies across domains. 

We focus on the following baselines and report per-participant metrics along with average metrics for both studies across all relevant domains.
\begin{itemize}
    \item "1-second Dense Captioning with LaViLa" - this baseline densely splits the video into 1-second long clips and uses LaViLa \cite{lavila} on each clip to caption individual moments in the video.
    \item "Ultrasonic Action Captioning Without Video" - this baseline uses models trained on the active acoustic sensing modality to generate clips based on whether the ultrasonic classifier (in this case a pre-trained ActSonic \cite{ActSonic} model) classifies a particular 1-second clip as within the domain. The caption for this domain is derived from the classifier label (e.g. for a particular label ``eating", the caption would be extracted as ``C performed the action: eating"). Notably, this method does not need to sample the video at all, but misses vital context which could be useful for understanding the details of the action.
    \item \name{}, using ultrasonic action detection (via a pre-trained ActSonic \cite{ActSonic} model) to efficiently clip a video before applying the LaViLa narrator to build an activity record.
    
\end{itemize}

\subsection{Quantitative Results}
We report per-participant metrics in Table \ref{table:us-results}. We find a relatively large reduction (avg 68\%, max 95.9\%, min 34.7\%) in activity records using active acoustic sensing with relevant domain actions, though reductions are uneven due to the uneven distribution of eating activities (e.g. P06 spent most of the session eating, resulting in a low reduction of the activity record). We found a higher alignment score by combining both ultrasonic and video modalities to capture and record activities when compared to only using the cheaper ultrasonic modality (0.892 avg for \name{} vs 0.828 avg for ActSonic, with low alignment values primarily due to a lack of relevant details within the corresponding activity documents, preventing the LLM from giving a detailed response). Notably, these high correlations and significant \% reductions are achieved without fine-tuning either the ultrasonic activity clipper or the visual captioning model on new videos, resulting in ``session-independent/user-independent" performance metrics. In addition, these results are collected on user study data that is primarily centered around eating activities: if extended to longer ``everyday recordings" where eating is comparatively sparse, future iterations of this system could achieve much higher record reduction metrics.
\begin{table}[h]
\begin{tabular}{l|l|l|l}
\textbf{} & \textbf{EchoGuide vs Dense} & \textbf{ActSonic vs Dense} & \textbf{\% Reduction} \\ \hline
P01                  & 0.888                                                 & 0.854                                                & 95.1\%                          \\
P02                  & 0.897                                                 & 0.823                                                & 53.4\%                          \\
P03                  & 0.873                                                 & 0.866                                                & 83.7\%                          \\
P04                  & 0.888                                                 & 0.862                                                & 95.5\%                          \\
P05                  & 0.906                                                 & 0.774                                                & 55.1\%                          \\
P06                  & 0.890                                                 & 0.805                                                & 34.7\%                          \\
P07                  & 0.882                                                 & 0.785                                                & 40\%                            \\
P08                  & 0.897                                                 & 0.853                                                & 95.9\%                          \\
P09                  & 0.907                                                 & 0.833                                                & 67.5\%                         
\end{tabular}
\caption{\name{} metrics across participants in user study, including correlation with dense 1-second clipping (measured as mean BERT F1 score across all sessions per participant) vs ActSonic correlation with 1=second clipping, and \% reduction in activity record using active acoustics vs 1-second clipping}.
\label{table:us-results}
\end{table}

\section{Evaluating activity records' ability to answer targeted eating questions with large image-language models}
\label{sec:eval2}

\begin{table}[h]
\begin{tabular}{|l|l|l|l}
\textbf{} & \textbf{Food type  (1fps/0.5fps)} & \textbf{Utensil type}  & \textbf{Container type}  \\ \hline
P01         & 0/0                              & 0/0                   & 1/0                     \\
P02         & 1/1                           & 1/1                   & 1/1                     \\
P03         & 0/0                              & 1/0                   & 1/0                     \\
P04         & 0/0                              & 0/0                   & 0/0                     \\
P05         & 0/0                              & 1/0                   & 0/0                     \\
P06         & 1/1                              & 1/1                   & 1/1                     \\
P07         & 1/0                              & 1/1                   & 1/1                     \\
P08         & 1/1                              & 1/1                   & 1/1                     \\
P09         & 0/0                              & 1/1                   & 1/1                    
\end{tabular}
\caption{Results of manual evaluation of \name{} + GPT4o given 1fps vs 0.5 fps sampling of frames from ActSonic-defined clips (based on zero-shot accuracy). Notation is defined as (X/Y) where X=accuracy at 1fps and Y=accuracy at 0.5fps}
\label{table:gpt4o-gt-man-evals}
\end{table}

\subsection{Metrics/Evaluation Procedure}
\name{}, however, focuses not only on providing general summaries via activity records of an individual's day from video and wearable sensor data, but also on answering targeted questions about these summaries by leveraging the image-text pertaining of large multimodal language models. We evaluate this method via manual review and annotation of the system's answers to three eating questions (``What did C eat/drink? What utensils did C use while eating/drinking? What container did C eat or drink out of?") when configured to use GPT4o \cite{GPT4o} to caption images sampled at two varying FPS levels (1fps and 0.5fps) from clips proposed by ActSonic, and report accuracy metrics showing whether \name{} extracts correct values for these questions as compared to manually-determined "ground truth" (taken by watching the reference video and determining which item is present): we've shown accuracy values given 1fps and 0.5fps sampling in Table \ref{table:gpt4o-gt-man-evals}. Accuracy values are defined as a 0/1 binary: 0 represents responses that do not overlap with the ground truth, while 1 represents responses that do completely overlap with the ground truth. 

\subsection{Quantitative Results and Discussion}
In general, we find that while 0.5fps results in a slow reduction in performance for some participants, we can attempt to leverage only a few frames along with metadata information (e.g. classifier outputs) from active acoustic sensors to output useful information, instead of having to process an entire video which could be full of redundant frames. For more ambiguous class types such as ``food type" (which may not be easily determinable from appearance alone), we find a relatively low average F1 score (44\% for 1fps and 30.5\% for 0.5fps) across all participants, whereas for more recognizable/distinctive class type such as utensil type and container type, we find a relatively high average F-1 score (77\% for utensils and containers for 1fps, 55\% for utensils and containers for 0.5 fps). We find a clear performance drop as FPS is reduced (from 0.55 with 1 fps to 0.47 with 0.5 FPS), due to increased sparsity of frames causing reductions in visual detail for the models. As vision-language models and prompting techniques continue to improve, we expect these numbers will become more accurate over time.

\section{Discussion}
\label{sec:discussion}{Further Reduction on video recording when deployed in the wild} Our study results showed that \name{} helped reduce video recordings by an average of 68\% without significantly impacting the quality of summarization for eating activities. However, we want to point out that this percentage of reduction will likely be significantly higher if the system is deployed for full-day recording. In the user study, we only asked participants to collect data for 40 minutes, including the meal. Consequently, the ratio of eating activities in our dataset is significantly higher than it would be in a full day of recording. Therefore, if ActSonic is used to only activate the camera during eating activities in a full-day recording, the data reduction will likely be significantly higher than 68\%. Additionally, the frame rate of recording can be further reduced to answer specific questions, saving energy and processing resources. We plan to explore these questions further in future works

\begin{figure}[h]
\centering
\includegraphics[width=\linewidth]{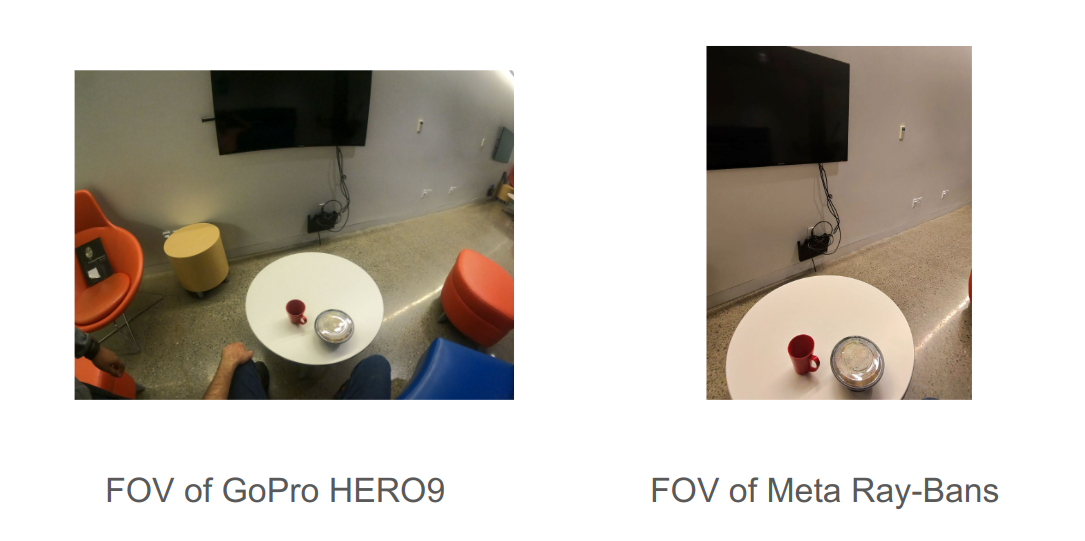}
\caption{Example showing field of view between GoPro vs Meta Ray-Bans}
\label{fig:side-by-side-fov}
\end{figure}
\textbf{Comparison to Egocentric Video recorded on glasses} The initial hardware system for \name{} was not collected using ``camera-enabled smartglasses" such as the Meta Ray-Bans \cite{MetaRayBans} due to limitations on recording videos with these off-the-shelf smart glasses. Instead, we used a head-mounted GoPro to easily capture egocentric activity videos. We found the information related to eating captured by smart glasses and our GoPro settings to be highly similar. To help readers understand the difference in images captured by these two devices, we used RayBan Smart glasses and a GoPro HERO9 mounted on the head to capture the same dietary scenario (a drink on the table), as shown in Figure \ref{fig:side-by-side-fov}. We found that the GoPro has a much wider field of view and can capture more general scene details in the orientation used by \name{} compared to the camera on the Meta RayBan smart glasses. However, because most foods are present near the center of the field of view, the difference in view angle between the two cameras did not impact the captured information. Therefore, the results reported in the paper can still be referenced for egocentric video analysis captured on smart glasses.


\textbf{Exploring additional domains for \name{}} While \name{} was adapted to focus primarily on localizing and understanding eating activities from video and acoustics, we also run a separate exploratory study with three participants operating in both the ``eating" and ``cooking" domains. Each participant engaged in 5 sessions of 4 minutes each within each domain with interventions between sessions to stop and restart data collection, for a total of 40 minutes per participant. 

The question for the "eating" domain was ``"What did the person C eat or drink?", with "relevant ultrasonic actions" being defined as the set of "eating", "drinking", and "pickup/putdown" (referring to manipulated items) and the question for the ``cooking" domain was ``What did the person C cook?" with relevant ultrasonic actions being the set of ``chopping", ``pouring", ``stirring", ``pickup/putdown", or ``walking". We configure LlamaIndex with GPT3.5-turbo and a temperature value of 0, as well as the standard context prompt ``You are a chatbot, able to have normal interactions, as well answer questions from the person about what they did today (walking, eating, cooking, etc). Here are the relevant documents for the context: \{context\_str\}. Instruction: Use the previous chat history, or the context above, to interact and help the user. Format responses as a paragraph."

We find a high average \% reduction of 87\% in record size across both domains by leveraging active acoustic sensing for clipping videos, along with higher correlation with dense captions (0.9 BERT F1 score) while using \name{}'s multimodal approach over only using active acoustic sensing (0.86 BERT F1 score). We find that combining the video and ultrasonic modalities additionally shows quantitative improvement (with respect to alignment with the dense caption summary of the original video) when compared to only using the ultrasonic modality, while still maintaining high reductions in the activity record. Though more thorough investigation needs to be done to show this system can work across a wider variety of everyday activities, improvements in unseen domains show the relatively task-agonstic nature of the \name{} software pipeline. 

\textbf{Improving comfort and reliability of hardware prototype} The current hardware prototype leverages a Teensy-based microcontroller on the left side of the eyeglasses which is connected to a phone for power and recording control, along with a Go-Pro head-mounted camera for video capture. The relative weight and complexity of the combined devices were cited in user surveys as the primary reason for the low comfort rating of the prototype (as it weighed more on the head and ears).

Leveraging a lower-power BLE with a LiPo battery (similar to Google Glass) as the primary microcontroller module for active acoustic recording, along with a Flex-PCB that reduces extraneous wiring, can reduce the unwieldy nature of the acoustic system. Developing custom low-power, high-FPS and high-resolution cameras (such as event-based cameras) that are purpose-built for eyeglass frames can also enable seamless video recording without a GoPro requirement, reducing the weight of the systems considerably. Building eyeglass frames that can swap lenses in a custom way, or building a system that can be seamlessly applied on any eyeglass, can reduce the likelihood of participants with custom prescriptions being unable to see through the provided eyeglasses. 

\textbf{Reducing software latency to enable real-time applications} Currently, \name{} processes and asks questions over activity records in an offline fashion, but many users may want to understand their activities in an online fashion (for instance, asking about previous meals while evaluating what food to get at a restaurant). As seen in other concurrent works \cite{echospeech}, active acoustic postprocessing could be completed on a smartphone, and with advancements in embedded AI chips and stronger networking modules for more robust cloud access, it may be possible to do end-to-end inference online with both edge-deployed and cloud-deployed models.

\textbf{Improving overall model flexibility to new situations} While Sec \ref{sec:eval} and Table \ref{table:us-results} show promising results for \name{} usage (combining video and ultrasonic modalities) across two distinct domains and procedural styles in everyday activities, further improvements can be made to enhance overall system performance. Collecting and fine-tuning on a larger base dataset of ultrasonic captures of activities can enable more robust, user-independent detection of human body motion, while leveraging steadily more powerful large multimodal models can enable more robust and generalizable video captions that encode more domain-specific or estoeric information. 

The \name{} pipeline leverages a Large Language Model with Retrieval Augmented Generation to enable document-based question-answering, along with a Large Multimodal Language Model to enable video captioning. Further optimization of system performance could be achieved via careful prompt engineering methods, such as chain-of-thought with few-shot exemplars \cite{Cot, ZsCot}. We leave this in-depth exploration of prompt engineering methods to future work.

\section{Conclusion}
\label{conclusion}

In this paper, we present \name{}, an innovative application pipeline that combines low-power active acoustic sensing on eyeglasses, egocentric video analysis, and large-scale language models to efficiently detect and analyze eating activities. Our evaluation with 9 participants in naturalistic settings demonstrates that \name{} achieves high-quality summarization with a significant reduction in record size while maintaining high semantic correlation with densely-captioned records. As smart glasses become more widespread and equipped with various sensors, multi-stage pipelines like \name{} have the potential to be applied to a broader range of activities and contexts without requiring explicit fine-tuning for individual users.

\section*{Acknowledgement}
This project was supported by the National Science Foundation Grant No. 2239569 and partially by the Cornell University IGNITE Innovation Acceleration Program. We also acknowledge and thank Prof. Bharath Hariharan at Cornell CS for help with computing resources.

\bibliographystyle{ACM-Reference-Format}
\balance
\bibliography{refs}

\end{document}